# Quantum holography with single-photon states

Denis Abramović [1,2,*], Nazif Demoli [1,2], Mario Stipčević [2], and Hrvoje Skenderović [1,2]

[1]Institute of Physics, Bijenicka 46, Zagreb 10000, Croatia

[2]Centre of Excellence for Advanced Materials and Sensing Devices, Photonics and Quantum Optics Unit, Rudjer Boskovic Institute, Bijenicka 54, Zagreb 10000, Croatia

*Corresponding author: *dabramovic@ifs.hr*

The retrieval of the phase with single-photon states is a fundamental and technical challenging endeavor. Here we report the first experimental realization of hologram recordings with heralded single-photon illumination and continuous observation of photon statistics. Thereby, we demonstrate the basic principle of holography with single-photon states which cannot be described with the classical wave theory. Under conditions with illumination more than 200 times weaker than the noise of the detector, a hologram (interferogram) recorded with a heralded single-photon source revealed an object not visible with non-heralded illumination and slightly higher intensity. The dramatic improvement in retrieval of amplitude and phase information achieved with the heralded single-photon source can be explained by the strong suppression of noise due to the nonclassical temporal correlation between twin photons and the small coincidence time window. The method could be useful for recording and retrieving of amplitude and phase information in the presence of strong noise, for covert imaging, and for imaging of photosensitive biological and material samples.

## I. INTRODUCTION

Holography is a powerful imaging technique that can record and retrieve both amplitude and phase information [1] across a broad electromagnetic spectrum. Besides classical techniques developed for recording holograms in low-light-level conditions with weak coherent state (attenuated laser) and photon counting [2-4], quantum concepts using down-conversion light source and complex beam preparation setups for recording holograms were also investigated. These quantum experiments are based on two-photon probability amplitude [5], polarization entanglement [6], one-photon probability with two object beams [7] and interference of two beams from two separate down-conversion processes [8]. Strong laser pumping of the down-conversion crystal for generating photon pairs can lead to multi-photon pair emission [9, 10] and losses in the system can lead to photon detections without true single-photon characteristics [11, 12] of the experiment. In experiments, often for practical purposes just an attenuated or not well characterized source is used that cannot fulfil criteria for a single-photon source. Apart from tackling different aspects and technical issues of hologram recording, none of these classical [2-4] and quantum [5-8] holographic methods characterize the nature of the light source with the second order correlation function, $g^{(2)}(0)$. This parameter is standardly used for characterization and verification of a single-photon source. In complex experiments where the single-photon character is of a fundamental interest, the source can be characterized either separately [12, 13], at certain points during the experiment [14] or during the full course of the experiment [11]. The characterization of the illumination during the full course of the experiment gives maximal experimental assurance that the source maintains the single-photon character and no additional hypothesis about the character of light source are necessary.

The retrieval of the phase from a hologram with single photons is a challenging task due to the indeterminate global phase of single photons [5, 7, 15] and can be potentially obscured by the nature of the recording. It may appear puzzling to record a hologram with number states, because for single-photon states or for any precisely defined Fock states the quantum phase cannot be measured simultaneously with an arbitrarily high precision [16]. Theoretically, the interference effects can be obtained due to the phase-dependent normal-mode expansion of the quantized electromagnetic field and its corresponding time-evolution [17]. However, no conclusive experimental evidence of a successful hologram recording with single-photon states was demonstrated until now.

In this paper, we present an approach for recording holograms in a basic holographic scheme with continuously monitored light source and single-photon detection. The implementation of the single-photon illumination and exclusion of the classical wave theory is supported via a long-run measurement of the second order correlation function, $g^{(2)}(0)$ separately and during the two-dimensional multi-channel detection, in front of the image-plane holographic setup. In this way, evidence for non-classical character of the experiment is established by the measured photon statistics. The comparison of simultaneously recorded holograms with heralded single-photon states and non-heralded light as well as their amplitude and phase reconstructions are shown. As an object for hologram recordings, we used a silver mirror with a laser written pattern. The details about fs laser parameters for writing can be found in Ref. [18]. The single photons are generated by the heralded single-photon source in the process of spontaneous parametric down conversion (SPDC). The single-photon sensitivity is achieved by the usage of single-photon counting module (SPCM) based on single-photon silicon avalanche photodiodes.

## II. THEORETICAL BACKGROUND

### A. Principle of holography with single-photon states

The basic principle of holography consists of two steps, the first is recording of an image that contains the amplitude and phase information, and the second step is a reconstruction of the amplitude and phase information. The information is coded by interference between the reference beam and the diffracted beam from the object. From quantum perspective, the coding process can be described as the quantum superposition of a single particle. The single particle (S) is associated with two probability amplitudes, $r$ and $o$, where $r$ is the amplitude of the single particle that goes over the reference mirror (M) to the detector (D) and $o$ is the amplitude of the single particle that goes over the object (O) to the detector. The information coding process can be described in Dirac notation by following basic equation of holography:

$$\langle D|S\rangle_{\text{via M or O}} = \frac{1}{\sqrt{2}}\left(\underbrace{\langle D|M\rangle\langle M|S\rangle}_{r} + \underbrace{\langle D|O\rangle\langle O|S\rangle}_{o}\right) \quad (1)$$

From Eq. (1), the probability that the photon reaches the detector (in the detection plane) is

$$P = \frac{1}{2}\left(|r|^2 + |o|^2 + 2|\text{r}||\text{o}|\cos(\varphi_r - \varphi_o)\right) \quad (2)$$

where $r = |\text{r}|\exp(-i\varphi_r)$ is probability amplitude associated with reference path and $o = |\text{o}|\exp(-i\varphi_o)$ is probability amplitude due to the object path. For holography, only the last term of the Eq. (2) is interesting, because this term contains the phase information. In the proposed image-plane holographic system with an off-axis beams, the phase terms in the Eq. (2) can be considered to contain the phase due to respective angle between the reference and object beam. The decoding process can be done by an illumination of the hologram with reference beam or as in this work by the numerical reconstruction. The numerical reconstruction of the image-plane holograms can be described by the following steps: (i) a 2-dimensional Fourier transform of the hologram is taken, (ii) in the frequency domain, the first diffraction order is spatially extracted, (iii) an inverse Fourier transform is taken on the spatially extracted image. After this procedure amplitude and phase image are obtained. However, to remove the linear phase term introduced by fundamental fringes due to off-axis beams, phase correction of the phase image (obtained in step (iii)) is done numerically. In this way a single hologram recording is sufficient to obtain the phase information. The mathematical procedure for the phase correction is described in Ref. [19].

The image-plane setup is realized as a modified Michelson interferometer (Twyman-Green interferometer), which uses in one of the arms a reflecting object instead of a mirror. Fig. 1 shows setups for characterization of light before the interferometer with detectors D2 and D3 and after the interferometer with detectors D4 and D5. By assuming lossless 50/50 beam splitter, non-vacuum state at the input port (1), and a vacuum state at the second input, the expectation value of the photon number operator at the interferometer output (I) is

$$\langle \hat{n}_I \rangle = \langle \hat{n}_1 \rangle \frac{1}{2}\left[1+\cos(\varphi_r - \varphi_o)\right], \tag{3}$$

where $\varphi_r$ and $\varphi_o$ are phase accumulations due to back reflected reference and object path, respectively.

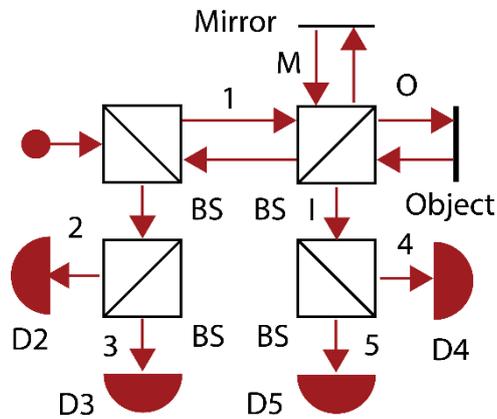

FIG 1. Conceptual scheme of modified Michelson interferometer with characterization setups. The arrows correspond to the possible path of single photons. The circle denotes a non-vacuum input state. Numbers (1-5) and letters (M, I, O) are associated with the shown ports. BS is 50/50 beam splitter.

The joint probability of detecting a photon at the detectors D2 and D3, or D4 and D5 for the single-photon state at the input port, and a vacuum state at the second input port is

$$P_C = \langle \hat{n}_2 \hat{n}_3 \rangle = \langle \hat{n}_4 \hat{n}_5 \rangle = 0. \tag{4}$$

The last relation is without analogy for a classical field [20] and can only be explained in terms of single-photon states. Based on this fact, we designed the experiment to verify the single-photon nature of our device and to demonstrate the feasibility of phase retrieval by recording a hologram with single-photon states. The Eq. (3) which describes the interferometer, corresponds to the Eq. (2), the basic equation of holography. The last term in these two equations, which includes the phase, is the only relevant term for the coding principle of holography. Note that both of these equations are interpreted from quantum perspective and only an input with single-photon state can satisfy these equations and Eq. (4) simultaneously.

### B. Model for contrast enhancement

Quantum imaging methods allow imaging that goes beyond conventional classical imaging. Roughly speaking, the advantages of quantum imaging can be divided into two classes: contrast or signal-to-noise enhancement [21, 22] and resolution enhancement [23, 24]. Most quantum imaging techniques rely on either quantum resources such as entanglement or spatial correlation. However, in a simple model that does not assume any complex resource such as entanglement or spatial correlations, England et al. [25] demonstrate (amplitude) image contrast enhancement using configurable background noise from an independent light source and nonclassical temporal correlations [26] between twin photons. Our experiment and model are also based on a comparison of the method with and without temporal correlations between twin photons but with some important differences. Among the most important features of our experiment is that we demonstrate contrast enhancement for both the amplitude and phase images (by comparing classical and quantum hologram) and that our model does not assume that the probability of registering a noise count is same for method based on temporal and without temporal correlation. In this way, we provide a more realistic model that accounts for the fact that coincidence noise is different from the noise registered in imaging without a temporally correlated technique by a single detector. Another difference is that our method does not require the acquisition of two separate images to quantify the experimental SNR. The latter is due to the fact that we separate the hologram image into two regions: the bright region covers an area of constructive interference from which a signal can be registered, while the dark region covers an area of destructive interference or an area from which no signal is expected. In the following subsections, we analyze two techniques: the heralded technique based on temporal correlations between twin photons and the non-heralded technique based on detection of events without temporal correlations.

#### 1. Heralded SNR

The following model assumes the generation of temporally correlated twin photons. Ideally, one of the twin photons is detected by the trigger detector (SPCM1 shown in Fig. 2) and the other signal photon by the imaging detector (SPCM4 or SPCM5 shown in Fig. 2). If we denote the probability of detecting coincidence signal as $P_{CS}$, the probability of detecting coincidence (originating either from signal or noise) as $P_C$, and the probability of detecting coincidence noise as $P_{CN}$ then $P_{CS} = P_C - P_{CN}$. All the probabilities are stated for detecting an event in a given time bin (coincidence time window). The signal-to-noise ratio (SNR) for heralded technique can therefore be defined as

$$SNR_H = \frac{P_{CS}}{P_{CN}} = \frac{P_C - P_{CN}}{P_{CN}} \qquad (5)$$

The probability of coincidence noise $P_{CN}$ is composed of probabilities related to the coincidences between the dark count of the imaging detector and trigger detector, $P_{DI-T}$, coincidences between the imaging detector and the dark counts of the trigger detector, $P_{I-DT}$, and coincidences between the dark imaging detector and the dark trigger detector, $P_{DI-DT}$. Using this notation, the probability of coincidence noise is given by

$$P_{CN} = P_{DI-T} + P_{I-DT} - P_{DI-DT} \qquad (6)$$

The minus term in the definition of $P_{CN}$ must generally be included, because the dark counts are already included twice in terms that contain $P_{DI-T}$ and $P_{I-DT}$.

For sake of relating probabilities to experimentally measured counts, it is useful to define the following mean variables: $N_{CS}$ as coincidence signal, $N_C$ as coincidence (originating either from signal or noise) and $N_{CN}$ as coincidence noise. Using this notation, $N_{CS} = N_C - N_{CN}$. All the stated variables can be considered as average counts per pixel during the integration time $T$ at the single pixel. Regarding the number of time bins $B$, it is necessary to define the coincidence time window $\Delta t$ within which the coincidences are selected, so that $B = T/\Delta t$. These variables are related to probabilities by following relations $P_{CS} = N_{CS}/B$, $P_C = N_C/B$ and $P_{CN} = N_{CN}/B$. The heralded SNR can therefore be written as

$$SNR_H = \frac{N_{CS}/B}{N_{CN}/B} = \frac{N_C - N_{CN}}{N_{CN}} \qquad (7)$$

Note that $SNR_H$ is always $\geq 0$, because $N_C$ includes $N_{CN}$ and $N_{CS}$. The coincidence noise ($N_{CN}$) is the coincidence noise related to the dark counts of the imaging detector $N_{DI}$, counts registered by the imaging detector $N_I$, the dark counts of the trigger detector $N_{DT}$, and counts registered by the trigger detector $N_T$. Thus, the coincidence noise can also be written in terms of average counts per pixel during the integration time at the single pixel as

$$N_{CN} = N_{DI} \cdot N_T \cdot \Delta t + N_I \cdot N_{DT} \cdot \Delta t - N_{DI} \cdot N_{DT} \cdot \Delta t \tag{8}$$

The minus term in the definition of $N_{CN}$ must be included, because the dark-dark coincidence noises are included in each of the first two terms on the right-hand side of Eq. (8).

## 2. Non-heralded SNR

If $P_S$ denotes the probability of detecting a photon (originating either from signal or noise) on the imaging detector (SPCM4 or SPCM5) and $P_{SN}$ is the probability of detecting a (singles) noise on the imaging detector then the probability of detecting a (singles) signal on the imaging detector $P_{SS} = P_S - P_{SN}$. SNR for non-heralded technique can be defined as follows

$$SNR_{NH} = \frac{P_{SS}}{P_{SN}} = \frac{P_S - P_{SN}}{P_{SN}} \tag{9}$$

By defining following mean variables, $N_{SS}$ as singles signal, $N_S$ as singles and $N_{SN}$ as singles noise, then $N_{SS} = N_S - N_{SN}$. As before, all variables are stated as average counts per pixel during the integration time $T$ at the single pixel. Furthermore, all probabilities are related to the maximum number of events $N$ that could be recorded without accounting for losses during the integration time $T$. In this way, $P_{SS} = N_{SS}/N$, $P_S = N_S/N$ and $P_{SN} = N_{SN}/N$. Thus, SNR for non-heralded technique can be written as

$$SNR_{NH} = \frac{N_{SS}/N}{N_{SN}/N} = \frac{N_S - N_{SN}}{N_{SN}} \tag{10}$$

If $N_S \approx N_{SN}$ or $N_{SS} \ll N_{SN}$ then $SNR_{NH}$ is approximately equal to zero. This is the case when the use of the heralded technique may be particularly beneficial because of the possible noise suppression.

### 3. Heralded enhancement factor

Finally, we can express heralded enhancement factor as follows

$$HEF = \frac{SNR_H}{SNR_{NH}} = \frac{P_{CS}/P_{CN}}{P_{SS}/P_{SN}} = \frac{P_{CS} \cdot P_{SN}}{P_{SS} \cdot P_{CN}} \qquad (11)$$

In addition, it is possible to introduce a factor $\eta$ that takes into account the efficiency that heralded event is included in non-heralded events. In this case $P_{CS} = \eta \cdot P_{SS}$, and heralded enhancement factor yields,

$$HEF = \frac{\eta \cdot P_{SS}/P_{CN}}{P_{SS}/P_{SN}} = \eta \frac{P_{SN}}{P_{CN}} = \eta \frac{N_{SN}}{N_{CN}} \qquad (12)$$

It is clear from Eq. (12) that the heralded enhancement factor is greater when the noise suppression is greater. The whole analysis can be applied to recorded holograms so that the signal consists of the photons recorded in areas of constructive interference. Since the detector cannot distinguish between noise and signal, we define a bright region as an area from which a signal and noise can be registered, while a dark region is defined as an area that is a consequence of destructive interference or an area from which no signal is expected.

In our work, we compared the results of the heralded and the non-heralded technique with an approximately equal number of signal photons. The typical coincidence time window for silicon single-photon avalanche detectors has a size of a few ns and is mainly limited by detection jitter. The small coincidence time window leads to a selection of events (coincidences) related to temporal correlations between twin photons. In other words, the imaging detector does not capture all the noise that is detected in the continuous mode of operation, but it captures only events when signal is expected. Therefore, the use of a small coincidence time window is a key factor in noise suppression. However, it should be noted that the noise level in the hologram (interferogram) may be limited by the visibility of the fundamental fringes. For example, the source of this noise can be caused by the ability to distinguish a photon that emerges either from reference or object path. This imposes the maximum possible visibility of interference fringes.

Anyway, remark that the previous analysis was made under the assumption that the photon is equally likely to come from either the reference or object arm of the interferometer.

### III. EXPERIMENTAL SETUP

As shown in Fig. 2, the experimental setup for single-photon holography, consists of three parts: a light source (a), a setup for continuous characterization of light before the interferometer (b), and an interferometric image-plane setup with a scanning single pixel detection and characterization system (c). The object used for the hologram recording is shown in part of Fig.2(c). In the heralded single-photon experimental scheme, the laser beam is weakly focused into the nonlinear crystal where photon pairs are generated. The coupling of the photon pairs into the single-mode fibers destroys the spatial correlations between the twin photons. Thermal states are naturally generated by considering single arm (non-heralded light) of the SPDC source. From the perspective of our experimental setup, the photon statistics for one arm is very similar to a Poissonian [27], because all relevant parameters for the photon statistics are averaged out due to resolving times much larger than the coherence time of the down-converted photons. Therefore, it can be considered that the thermal states generated in the SPDC are practically imitating coherent states in terms of photon statistics.

The hologram image is obtained by on-the-fly and line-by-line scanning, with line acquisition always starting on the same side of the raster scan path (illustrated in Fig.2). The pixel size in the x dimension is equal to the product of the stage velocity ($v_x$) and the integration time ($\Delta\tau$), $\Delta x = v_x \cdot \Delta\tau$. The pixel size in the y dimension is made the same size by moving each consecutive line by $\Delta y = \Delta x$. All the detection channels are connected to time tagging module (TTM). TTM streams the counts from real input channels and the software-defined channels for all relevant double and triple coincidences over USB to PC storage simultaneously. This allows simultaneous recording of holograms or photon statistics with heralded and non-heralded light.

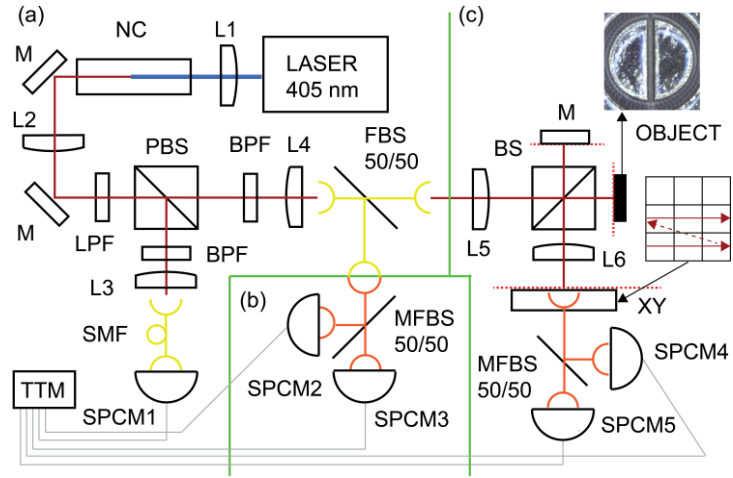

FIG 2. Experimental setup for recording holograms with single-photon states and non-heralded light. In part (a) the single frequency (volume Bragg grating stabilized) diode laser with 8 mW illuminates 10 mm long ppKTP nonlinear crystal (NC) to produce collinear, orthogonally polarized, and degenerate photon pairs at 810 nm. The laser beam passes the focusing lens (L1) and is blocked by long pass filter (LPF). The down-converted photons are collimated by lens (L2) and further filtered from background light by band pass filter (BPF) with central wavelength of 810 nm. The vertically polarized photons reflected from the polarizing beam splitter (PBS) are focused and coupled into single-mode fiber through aspheric lens (L3) and detected by the detector SPCM1. The second, horizontally polarized photon is guided through aspheric lens (L4) into single-mode fiber that consist of a non-polarizing beam splitter (FBS) with 50% transmission and 50% reflection (50/50). The reflected photons enter the part (b) where multimode fiber beam splitter (MFBS 50/50) guides the light to SPCM2 and SPCM3. The transmitted photons enter part (c) where the light from single-mode fiber is collimated by lens (L5) and directed towards 50/50 beam splitter (BS), mirror (M) and object. Finally, the imaging lens (L6) images the object (and the mirror) plane to the detector plane (dashed lines). In the detection plane, a single end of the (second) multimode fiber beam splitter (MFBS 50/50) is attached to the two-dimensional motorized linear translation stage (XY). The other end of the fiber feeds the light into SPCM4 and SPCM5.

Optimal matching between the length of the two optical arms of the interferometer was found by movement of the reference mirror with linear motorized translation stage to the position with highest visibility. The measurement was performed with a bucket detector instead of the "pixel" detector (shown in Fig. 2) behind the interferometer. In the next step, the bucket detector was removed, and the interference fringes were aligned so that the imaged object was properly taken into account. In this way, an optimal visibility of the fringes can be obtained for any kind of light source with stable coherence length.

## IV. RESULTS

### A. Photon statistics related to nature of light

The timing correlation between heralding and heralded single photon represents essential part of single-photon generation with the SPDC process. Even though necessary, it is not sufficient evidence for an exclusion of classical wave theory [11, 12, 14]. To rule out the classical wave theory, we experimentally evaluate the second order correlation function $g^{(2)}(0)$ for three detectors, heralding detector and two heralded detectors placed behind the beam splitter. We use the experimental definition [14] for which

$$g^{(2)}(0) = (N_1 \cdot N_{123}) / (N_{12} \cdot N_{13}) \tag{13}$$

where $N_1$ (SPCM1) is number of single events registered by heralding detector, $N_{12}$ is the number of double coincidences registered between heralding (SPCM1) and heralded detector (SPCM2), $N_{13}$ is the number of double coincidences registered between heralding (SPCM1) and second heralded detector (SPCM3), and $N_{123}$ is number of triple coincidences registered between heralding (SPCM1) and two heralded detectors (SPCM2 and SPCM3) behind 50/50 beam splitter. The value $g^2(0) \geq 1$ holds for classical wave theory [11, 12, 14] and when $g^2(0) \leq 0.5$ is achieved, then the light source is considered a good single-photon source due to the nonzero projection on the single-photon Fock state, where value closer to zero indicates a purer single-photon state [28].

For demonstration of photon statistics of marginal (one) SPDC arm [27], we use (normalized) second order correlation function

$$g^{(2)}(0)_{nh} = (N_{23} \cdot T_T) / (N_2 \cdot N_3 \cdot \Delta t) \tag{14}$$

where $N_{23}$ is the number of double coincidences (SPCM2 and SPCM3), $N_2$ (SPCM2) and $N_3$ (SPCM3) are the number of single events registered behind the two detectors behind the beam splitter, $T_T$ is total time of measurement and $\Delta t$ is the size of coincidence time window. Photon statistics of marginal SPDC arm and of the two SPDC arms can be recorded from same SPDC source simultaneously.

Before recording holograms, we characterized $g^{(2)}(0)$ using SPCM1 as heralding detector and dielectric mirror instead of the first FBS 50/50, so that all the light goes just to SPCM2 and SPCM3 which are now heralded detectors. In Fig. 3 (upper inset), the measurement of the second order correlation function clearly demonstrates single-photon character of the implemented light source through 24 hours. The final value of the second order correlation function, $g^{(2)}(0) = 0.00440(1)$ is among the best recorded compared to the literature [29]. The $g^{(2)}(0)$ error is small due to accumulation of events over a long period of time and is determined by applying the Poisson distribution to the count rates. Measurement of the second order correlation function for a single SPDC arm gives $g^{(2)}(0)_{nh} = 1.00115(107)$ and it is also shown in Fig. 3 (lower inset). This result agrees with the literature [27] and illustrates the difference between the photon statistics associated with heralded single-photon states and non-heralded light from SPDC. These two different natures of light, obtained from same SPDC source have been used for simultaneous recording of two holograms with continuous monitoring of photon statistics.

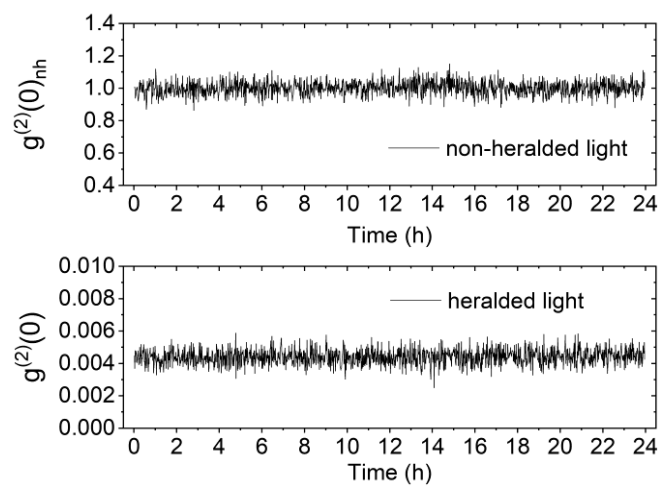

FIG 3. $g^{(2)}(0)$ and $g^{(2)}(0)_{nh}$ measured through 24 hours. Black line represents value of $g^{(2)}(0)$ and $g^{(2)}(0)_{nh}$ for measurements collected every minute. Note the different vertical scale on the two graphs.

When recording holograms in the setup shown in Fig. 2, the light source is monitored before the interferometer by SPCM2 and SPCM3 and after the interferometer by imaging detectors SPCM4 and SPCM5. The measurement of $g^{(2)}(0)$ with SPCM2 and SPCM3 during the recording of hologram shown in Fig. 4 gives $g^{(2)}(0) = 0.00721(3)$. The difference between the two observed values of $g^{(2)}(0)$ is due to losses introduced by additional fiber beam splitter and possible small alignment changes in the pump source during the two different experimental trials. The triple coincidence between SPCM1, SPCM4, and SPCM5 is zero due to the low photon count rates at SPCM4 and SPCM5. The photon count rates are low because the photons are spread over the detection image plane. Because of these low count rates, $g^{(2)}(0)$ cannot be explicitly calculated, but the zero value of the triple coincidences clearly indicates a single-photon behavior also at the image detection plane. These results of $g^{(2)}(0)$ measurement are fully consistent with the proposed single-photon generation and detection protocol for hologram recordings.

## B. Hologram recordings and reconstructions

Simultaneously recorded holograms and their reconstructions are shown in Fig. 4. In the hologram recorded with single-photon states, the two half-circles are clearly distinguished from the non-reflective background. However, the object is barely visible from the hologram recorded with non-heralded photons. The amplitude and phase reconstruction in the case of single-photon states, Fig. 4(b) and (c) shows a clear advantage over non-heralded light, Fig. 4(e) and (f). The recordings are 93x85 pixels and the integration time is 5 s/pixel. All holograms are recorded with a pixel size of 30 µm. The coincidence time window for hologram recorded with single-photon states is 2 ns.

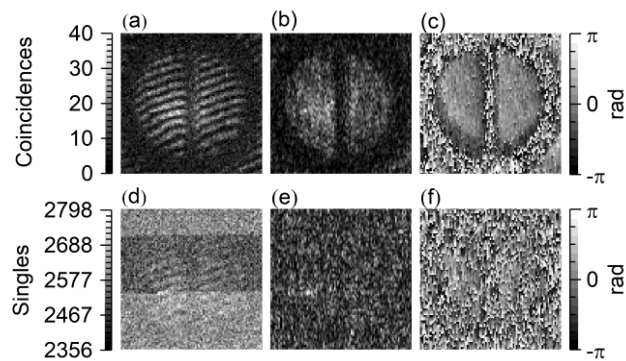

FIG 4. Upper row: hologram recorded with single-photon states (a); corresponding amplitude (b); and phase (corrected) reconstruction (c). Lower row: hologram recorded with non-heralded light (d); corresponding amplitude (e); and phase (corrected) reconstruction (f). Coincidence noise and noise of the imaging detector is not subtracted from the shown counts.

### 1. Visibility

Quantitatively, the quality of hologram fringes is derived from standard definition of visibility,

$$V = (N_{MAX} - N_{MIN}) / (N_{MAX} + N_{MIN}) \qquad (15)$$

where $N_{MAX}$ is local maximum and $N_{MIN}$ is next local minimum to maximum. For evaluation of the visibility, a line near the beam centre of the hologram recorded with heralded and non-heralded light was chosen. The profile of the line is shown in Fig. 5. The line corresponds to the vertical line taken on the horizontal pixel number 39.

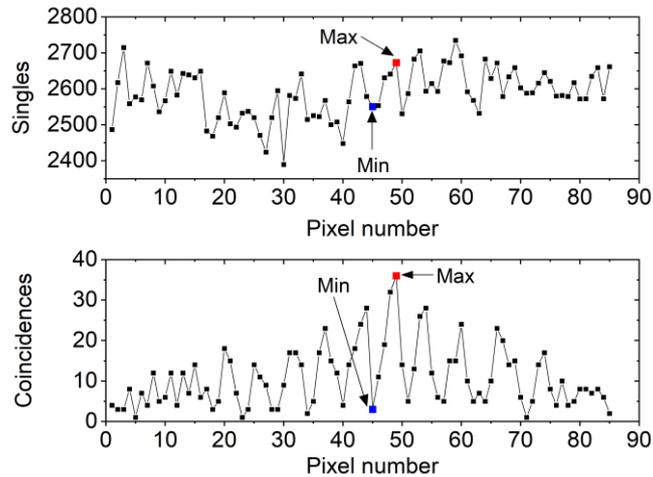

FIG. 5. Profile of single line taken from holograms recorded with heralded light (coincidences) and with non-heralded light (singles). Experimental data used to quantify local visibility are designated with (Min), denoting selected local minimum and (Max) denoting the selected local maximum.

Visibility near the beam center of the hologram recorded with non-heralded light is 2(1)%, and with single-photon states is 85(8)%. The complete results used to calculate visibility of the fringes are shown in Table I. The hologram recorded with non-heralded photons has low visibility due to predominant noise of the detector in comparison to signal from the interferometer. The dark noise of the detector is 2476(63) counts / (pixel · 5 s), and it agrees with the (local) minimum that is shown in Table I. Overall accidental coincidence noise calculated according to Eq. (8) for the later shown hologram in Fig.4(a) is 5(2) counts / (pixel · 5 s), and it is also in agreement with the (local) minimum. Spatial variations of visibility are predominantly due to properties of the object and the small, non-perfect and non-uniform Gaussian beam profiles of two overlapping beams that form the fundamental interference fringes, or, in other words, due to the decrease of intensity and spreading of illumination from the beam center. Depolarization of the beam entering the interferometer as well as depolarization effects inside the interferometer, for example due to the object may also influence the visibility. As can be seen in Table I, the maximum should be increased to 31672 counts to achieve the same visibility as for the heralded single-photon states without changing the count rate at the minimum. In other words, the maximum should be increased 879 times more than the registered heralded single-photon states.

TABLE I. Average counts / (pixel· 5 s) for a line in recorded holograms (Fig. 4(a) and (d)) and their visibilities.

|  | Single-photon states | Non-heralded light |
|---|---|---|
| Maximum | 36(6) | 2673(52) |
| Minimum | 3(2) | 2551(51) |
| Visibility | 85(8)% | 2(1)% |

**2. SNR and heralded enhancement factor**

For a more quantitative description of the overall quality of the hologram, we calculated and compared the directly measured and estimated signal-to-noise ratios according to the model presented in Section II.B. In summary, directly measured SNR is calculated according to Eq. (7) for the heralded technique and according to the Eq. (10) for the non-heralded technique. We do not consider the background noise separately from the dark noise, because detectors cannot distinguish between these two sources of noises, and we took special care to remove possible background noise due to stray light.

The dark region of the hologram recorded with single-photon states consists of pixels with the counts ≤7 and the rest of the pixels (counts >7) are defined as the bright region of the hologram. This threshold value was chosen because it agrees well with the estimated (average) coincidence noise and separates well the bright region from the dark region. The bright and dark regions used for both heralded and non-heralded case are depicted in Fig. 6. Due to much lower noise in the hologram recorded with heralded photons compared to non-heralded, we determine whether the pixel belongs to the bright or dark region from the hologram recorded with heralded single-photon states. The estimated coincidence noise can be calculated from the bright region using Eq. (8), where $N_I = 2579(65)$ counts / (pixel · 5 s), $N_T = 955018(7879)$ counts / (pixel · 5 s), $N_{DI} = 2476(65)$ counts / (pixel · 5 s), $N_{DT} = 2935(56)$ counts / (pixel · 5 s) and the coincidence time window $\Delta t = 2$ ns. From this is visible that the last two terms cancel, because $N_I \approx N_{DI}$. The data variability is stated as standard deviation.

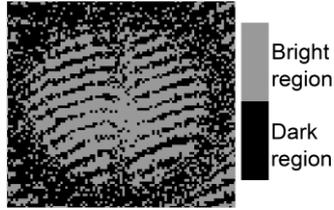

FIG. 6. Separation of bright and dark regions in recorded hologram.

The better contrast of holograms recorded with single-photon states is obvious from Fig. 4 and quantitatively from Table II. Table II shows excellent agreement between the measured and estimated SNR for the recordings with single-photon states. In the case of recordings with non-heralded light, the measured and estimated SNR show some difference related to an additional small number of background counts (approximately 4% of the detected events) compared to the estimated noise based on the measurement of dark counts of the imaging detector. In any case, the measured and estimated SNR for non-heralded case is approximately equal to zero. The standard deviations are relatively large, because of the fluctuations of dark counts of the SPCMs, some fluctuations in the intensity of the light source and non-uniformity of the Gaussian beam. The average heralded signal (coming from either the reference or the object arm) is 8 counts / (pixel · 5 s) and 11 counts / (pixel · 5 s) for the non-heralded signal. Despite the comparable signal level, the object is not visible with non-heralded light because the measured noise at the single-photon detector (2568 counts / (pixel · 5 s)) is more than 233 times larger than the non-heralded signal (11 counts/ (pixel · 5 s)). The visibility of the object for the non-heralded hologram can be easily lost due to the noise fluctuations of the imaging detector, which are much larger than the signal. Therefore, as is visible in Table II for the non-heralded light, the standard deviation of the signal with subtracted measured noise (92) is larger than the mean value (11) for non-heralded light. In comparison to the non-heralded hologram, the results show that nanoseconds coincidence time window used for recording holograms with heralded single photons leads to a much better contrast. The improvement is related to the strong suppression of noise, from 2568 to 5 counts / (pixel · 5 s). The directly measured coincidence noise in holograms recorded with heralded single-photon illumination consists mainly of the coincidence events detected in the small coincidence time window which is used for simultaneous detection of trigger events (SPCM1) and the noise of the imaging detector (SPCM4 or SPCM5). Other sources of noise are same as discussed regarding the sources of noises in the context of visibility of fringes (see Section 4.B.1). Clearly, non-heralded holograms are more susceptible (especially for long acquisitions) to any changes (see for example the dark stripe in Fig 4.(d)) such as light source intensity fluctuations, detector noise, fluctuations of detector noise, and residual background light, due to the continuous (non-heralded) acquisition of signal. If the goal were to achieve the same SNR with non-heralded light, then one would need to have 31672 detected imaging photons per pixel (in 5 s), or in other words more than 2879 times stronger illumination (by assuming the same detection noise and detection efficiency). Finally, in terms of heralded enhancement factor (*HEF*), we achieved enhancement of more than 500.

TABLE II. Average counts / (pixel · 5 s) for the three recorded holograms (Fig. 4(a) and (d)) and their SNRs. Estimated coincidence noise shown in the second column is obtained from bright region and estimated noise shown in the third column is obtained from dark noise of detector.

| Hologram | Single-photon states | Non-heralded light |
|---|---|---|
| Measured counts in bright region | 13(5) | 2579(65) |
| Measured counts in dark region | 5(2) | 2568(65) |
| Estimated noise | 5(2) | 2476(63) |
| Signal | 8(5) | 11(92) |
| Measured SNR | 2(1) | 0.004(35) |
| Estimated SNR | 2(1) | 0.04(4)) |

## V. CONCLUSIONS

In conclusion, we showed quantitative data about the photon illumination statistics to support the description of hologram recording from the framework of a single photon. Continuous measurement of photon statistics during hologram recording with single-photon states excludes with very high probability the possibility of a description based on classical wave theory. Compared to non-heralded light, our measurements with heralded single-photon states show strong improvement of contrast on both amplitude and phase reconstruction. This advantage of the coincidence technique can be useful for recording holograms in low-light conditions, in the presence of strong background noise, as well as for avoiding the use of expensive detectors with low dark count rates. Furthermore, the Michelson interferometer used is easily tuned to various off-axis configurations, and with the numerical method used, a single hologram image is sufficient to provide both amplitude and phase information. Unlike the method used to image objects in front of a Mach-Zehnder interferometer [7], the holograms are recorded without prior knowledge of the object-specific spatial frequencies. Note also that it is often considered that pixel-by-pixel imaging gives only incoherent image information [30], but our heralded single-photon experiment shows that imaging with single-pixel scanning preserves the amplitude and phase information at the same time.

A more ideal experiment with larger number of single-photon states and better (with higher efficiency and lower dark count rates) photon-number-resolving detectors behind the interferometer would be necessary for fully conclusive evidence for example against the argument that the light behaves differently in the characterization setup before the interferometer than behind the interferometer. However, the results of our experiment with the implemented continuous monitoring of the heralded single-photon states in front of the interferometer strongly indicate the non-classical nature of the illumination and the non-classical origin of the hologram.

This work was supported by the NATO SPS MYP G5618 and Ministry of Science and Education of Republic of Croatia grant No. KK.01.1.1.01.0001. D.A. acknowledges discussions with Marko Grba and Sören Wengerowsky.

# Supplemental Material: Quantum Holography with Single-photon States

**Description of the phase reconstruction procedure**

The employed numerical procedure to reconstruct image-plane holograms can be described as: (i) The hologram is obtained by scanning the fiber in the XY image plane of the interferometer. Two-dimensional (2D) data are then Fourier transformed yielding three separated terms, zeroth, plus first, and minus first order, in the frequency plane, (ii) an area around plus first diffraction term is isolated and all other points in frequency plane are set to zero, (iii) the resulting image with the isolated part is inversely Fourier transformed and a 2D complex function, $A(x,y)\exp[i\varphi(x,y)]$ is obtained. Here, $A(x, y)$ represents amplitude reconstruction and $\varphi(x,y)$ represents phase reconstruction. This phase includes a linear phase, $\varphi_r(x,y)$ introduced by off-axis configuration, and an object phase, $\varphi_o(x,y)$ which we are looking for, $\varphi(x,y) = \varphi_r(x,y) + \varphi_o(x,y)$. The linear phase can be removed by following: (iv) in the 2D Fourier plane from (i), a very small area in the center of the first order is taken (usually just 2×2 pixel), and inversely Fourier transformed to obtain $B(x,y)\exp[i\varphi_r(x,y)]$. Now, by taking complex conjugate and multiplying:

$$A(x,y)\exp[i\varphi(x,y)]B^*(x,y)\exp[-i\varphi_r(x,y)] = \\ A(x,y)B^*(x,y)\exp[i\varphi_o(x,y)]$$

the object phase can be extracted. The linear phase could be removed differently, for example, by moving the isolated area from (ii) to the center of the Fourier plane and then inversely transform; or by taking the hologram without the object to get $\varphi_r(x,y)$ directly.